\documentstyle[12pt]{article}
\begin{document}

\vspace*{6cm}

\begin{center}
{\bf{A note on closed string}}
\end{center}

\begin{center}
Shervgi S. Shahverdiyev$^*$
\end{center}

\begin{center}

{\small  Physical Institute, Azerbaijan National Academy of Sciences Baku, Azerbaijan}
 \end{center}
\vspace{.5cm}

\begin{quote}
\noindent
Special kind of closed strings is considered. It is shown that these closed strings
behave as two (an even number of) open strings at the classical level and one open string
at the quantum level. They contain massless vector field in their spectrum and can lie on D branes.
Some properties of closed string field effective action are declared.
\end{quote}

\vspace{1cm}

\vfill

e-mail:shervgi5@yahoo.com

www.geocities.com/shervgis

\newpage
\section{Introduction}
In this paper two problems in string theory are considered.
The first one is the presence of open strings in the type II closed string theories.
The second is the formulation of effective action for closed string fields.

With the appearance of D -- Branes \cite{p} in string theory, open strings ending on them were added
to type II theories. Natural appearance of open strings in the theory of closed strings remains unclear, to my best knowledge.

More progress has been done in understanding the role of open string tachyon field both in the framework of Cubic \cite{1} and Boundary \cite{2}-\cite{4} Open String Field theories (see \cite{6}-\cite{Tsey} and references therein). However, we do not know much about effective action for closed string fields \cite{b}.

In the present paper, we consider a subset of closed strings and show that they  behave as two (an even number of) open strings at the classical level. After quantization these closed strings contain massless vector field in their spectrum.
Massive fields in their spectrum have masses twice as much as the masses of fields in the spectrum of an open string. Therefore these closed strings behave almost as one open string at the quantum level. We apply this result to the effective action for closed string fields.

In the next section  we give a new look to closed strings and show that there are closed strings
that behave as two open strings at the classical level. Then we quantize these strings and show that they  contain massless vector field in their  spectrum and behave almost as one open string at the quantum level.
In section 3, we investigate properties of closed string spectrum and find some properties of closed string field effective action. Summary of results is given in section 4.
In section 5 we make remarks and give solutions that represent closed strings behaving as an even number of open strings at the classical level and almost one open string at the quantum level.

\section{Closed-open string duality}
Action for string
is
\begin{equation}\label{a}
S =-\frac{1}{4\pi\alpha}\int d \tau d \sigma
(-\gamma)^{1/2}\gamma^{ab}\partial_a X^\mu\partial_b X^\nu
\eta_{\mu\nu}
\end{equation}
and
$$
\delta S =
\frac{1}{2\pi\alpha}\int d \tau d \sigma \partial_a \{
(-\gamma)^{1/2}\gamma^{ab}\partial_b X_\mu\}\delta X^\mu -
\frac{1}{2\pi\alpha}\int d \tau \partial_{\sigma} X_\mu\ \delta X^\mu|_{\sigma=0}^{\sigma=\pi}
$$
so we have the following boundary condition \cite{GSW}-\cite{12}
\begin{equation}\label{*}
\partial_{\sigma} X_\mu\ \delta X^\mu(\tau, 0)-
\partial_{\sigma} X_\mu\ \delta X^\mu(\tau, \pi) =0.
\end{equation}
One of the solutions to this equation for closed string is
\begin{equation}\label{b}
\partial_{\sigma} X^\mu(\tau, 0)=\partial_{\sigma}X^\mu(\tau, \pi),\quad
X^{\mu}(\tau, 0)=X^{\mu}(\tau,\pi).
\end{equation}

In this paper we investigate properties of a special solutions to the equation of motion with boundary condition (\ref{b})
in conformal gauge represented by

\begin {equation}\label{1}
X^{\mu}=x^{
\mu}+ 2\alpha^{\prime}p^{\mu}\tau+ i(2\alpha^\prime)^{1/2}\sum_{n \ne 0}
\frac{1}{n}\alpha^{\mu}_n e^{-2in\tau}cos(2n\sigma).
\end {equation}
This solution has a few interesting properties as demonstrate in the sequel.
They satisfy the condition
$$
 \partial_\sigma X^{\mu}
(\tau,\pi/2)=0.
$$
Note that this condition is satisfied at point $\sigma =\pi/2$ and (\ref{1}) also satisfies
$$
X^{\mu}(\tau, 0)=X^{\mu}(\tau,\pi)
$$
which means that string is closed and behaves as two open strings because Neumann boundary conditions are satisfied in two different
points $\sigma=0$ and $\sigma=\pi/2$.

Next, we  quantize  closed strings represented by (\ref{1}).
To quantize (\ref{1}) we apply canonical quantization method, i.e., we impose
$$
[X^\mu(\tau, \sigma), \Pi^\nu(\tau, \sigma^\prime)]=i\eta^{\mu\nu}\delta(\sigma-\sigma^\prime),\quad
\Pi^\nu=\frac{1}{2\pi\alpha^\prime}\partial_\tau{X}^\nu
$$
which leads to the following commutation relations on the oscillators and the zero modes
$$
[x^\mu, p^\nu]=2i\eta^{\mu\nu},\quad [\alpha_n^\mu, \alpha _m^\nu]=n\delta_{n+m}\eta^{\mu\nu}.
$$
From  equation
$$
(L_0-1)|\psi_{phys}>=0, \quad
L_0=\frac{\alpha^\prime}{4}p^2+\sum_{n=1}^{\infty}
\alpha_{-n}\alpha_{n}+ \frac{D}{2}\sum_{n=1}^{\infty}n
$$
we can read the expression for the mass operator
$$
{\bf M}^2=\frac{4}{\alpha^\prime}\left(\sum_{n=1}^{\infty}\alpha_{-n}%
\alpha_{n} -1\right),
$$
As it is seen closed strings (\ref{1}) contain massless vector field $ A_\mu\sim \alpha_{-1}^\mu|0,k>$ in their spectrum.  We see that they behave as two open strings at the classical level and almost one at the quantum level, because the masses of fields are as twice as much as the masses in the spectrum of an open string.

\section{On closed string field effective action}

In this section, we assume that closed strings represented by
\begin{equation}\label{d}
X^{\prime\mu}=x^{\mu }+2\alpha ^{\prime }p^{\mu }\tau +i(\alpha ^{\prime}/2)^{1/2}
\sum_{n\neq 0}\frac{1}{n}\left( \beta _{n}^{\mu }e^{2in(\sigma
-\tau )}+{\tilde{\beta}}_{n}^{\mu }e^{-2in(\sigma +\tau )}\right).
\end{equation}
can transform to closed strings (\ref{1}).
Let us show that this corresponds to process ${\tilde{\beta}}_{n}^{\mu }\to \beta _{n}^{\mu }$.
The massless spectrum of closed strings (\ref{d}) is
$$
G_{\mu\nu} \sim (\beta_{-1}^\mu\tilde{\beta}_{-1}^\nu+\beta_{-1}^\nu
\tilde{\beta}_{-1}^\mu)|0,k> , \quad M^2(G_{\mu\nu})=0,
$$
$$
B_{\mu\nu} \sim (\beta_{-1}^\mu\tilde{\beta}_{-1}^\nu-\beta_{-1}^\nu%
\tilde{\beta}_{-1}^\mu)|0,k> , \quad M^2(B_{\mu\nu})=0,
$$
$$
\Phi \sim \beta_{-1}^\mu\tilde{\beta}_{-1}^\mu|0,k> , \quad M^2(\Phi)=0.
$$
After the limit ${\tilde{\beta}}_{n}^{\mu }\to \beta _{n}^{\mu }$ the field content changes. Fields $G_{\mu\nu}$
and $\Phi $ became massive and new field $ A_\mu$ appears.
$$
M^2=\frac{2}{\alpha^\prime}\left(\sum_{n=1}^{\infty}\left(
\beta_{-n}\beta_{n}+ \tilde{\beta}_{-n}\tilde{\beta}_{n}\right)-2\right)
\to {\bf M}^2=\frac{4}{\alpha^\prime}\left(\sum_{n=1}^{\infty}\beta_{-n}%
\beta_{n} -1\right),
$$
$$
G_{\mu\nu} \to {\bf G_{\mu\nu}} \sim \beta_{-1}^\mu\beta_{-1}^\nu|0,k> , \quad {\bf M}^2({\bf G_{\mu\nu})}= \frac{4}{\alpha^\prime},
$$
$$
B_{\mu\nu} \to 0,
$$
$$
\Phi \to {\bf \Phi} \sim \beta_{-1}^\mu\beta_{-1}^\mu|0,k>, \quad {\bf M}^2({\bf \Phi)}= \frac{4}{\alpha^\prime},
$$
$$
A_\mu\sim \beta_{-1}^\mu|0,k>, \quad {\bf M}^2(A_\mu)=0.
$$
We see that as ${\tilde{\beta}}_{n}^{\mu }\to \beta _{n}^{\mu }$ spectrum of closed strings (\ref{d}) coincides with the spectrum of
closed strings (\ref{1}).
According to this result we can state that the
limit ${\tilde{\beta}}_{n}^{\mu }\to \beta _{n}^{\mu }$ corresponds, in the language of fields,  to
\begin{equation}\label{eq}
S (G_{\mu\nu}, B_{\mu\nu}, \Phi, ...)\to {\bf S} ({\bf G_{\mu\nu}},
A_\mu, {\bf {\Phi}}, ...),
\end{equation}
where $S$ is the closed string field action.
This means that there is mechanism that can happen, and then the massless fields became massive and
new  massless vector field appears.

\section{Summary of results}

The main idea in this paper is that  open strings attached to D--Branes in type II theories do not satisfy the condition
$
X^{\mu}(\tau, 0)=X^{\mu}(\tau,\pi).
$
This condition should be satisfied because it is assumed from the beginning of the type II theories.
We suggest the way of avoiding this contradiction.

Closed strings  (\ref{1})

{{\bf{i)}} behave as two open strings at the classical level  because they satisfy
open string boundary conditions in two different points $\sigma=0$ and $\sigma =\pi/2$,

{\bf{ii)}} can lie on D-branes, because  after T duality on i-th direction  Neumann boundary condition becomes Dirichlet boundary condition on i-th direction. Therefore, two points of closed strings (\ref{1}),
$X^i_{closed}(\tau, \sigma=0)$ and $X^i_{closed}(\tau, \sigma=\pi/2)$,
 can lie on D--Branes,

{\bf{iii)}}} contain massless vector field in their spectrum and behave  as one open string at the quantum level.

The next result is  the following: if there is transformation between closed strings (\ref{d})
and closed strings (\ref{1}), this corresponds to the limit
${\tilde{\beta}}_{n}^{\mu }\to \beta _{n}^{\mu }$ in (\ref{d}),
then the closed string field effective action has  property (\ref{eq}).

\section{Remarks}
It is known that two or more open strings can join to form a
closed string. However, there is no evidence that the resulting
closed string will behave as two or more open strings. Moreover,
even and odd number of open strings can join to form a closed
string. In contrast, there are solutions to the equation of motion
with
\begin {equation}\label{y}
X^{\mu}_{closed}=x^{
\mu}+ 2\alpha^{\prime}p^{\mu}\tau+ i(2\alpha^\prime)^{1/2}\sum_{n \ne 0}
\frac{1}{n}\alpha^{\mu}_n e^{-2kin\tau}cos(2kn\sigma),\quad k\subset N
\end {equation}
satisfying Neumann boundary condition in 2k points
$$
\sigma=(0, \pi/2k, \pi/k, 3\pi/2k, ..., \pi).
$$
Hence, (\ref{y}) represents closed string behaving as 2k number of open strings. Therefore, there are no closed strings
behaving as an odd number of open strings. (\ref{1}) is (\ref{y}) with $k=1$.

We note that (\ref{1}) can be considered as a solution to the equation of motion with boundary conditions
$$
\partial_{\sigma} X^\mu(\tau, 0)=\partial_{\sigma}X^\mu(\tau, \pi) =0,\quad
X^{\mu}(\tau, 0)=X^{\mu}(\tau,\pi).
$$
These boundary conditions are also solutions to (\ref{*}).

\end{document}